\documentclass[aps,pra,floatfix,showpacs,noshowkeys,epsfig,graphics]{revtex4}%
\usepackage{graphicx}
\usepackage{amsmath}
\usepackage{amsfonts}
\usepackage{amssymb}

\newcommand{\newc}{\newcommand}
\newc{\beq}    {\begin{equation}}
\newc{\eeq}    {\end{equation}}
\newc{\beqa}    {\begin{eqnarray}}
\newc{\eeqa}    {\end{eqnarray}}
\newc{\ba}    {\begin{array}}
\newc{\ea}    {\end{array}}
\newc{\st}    {\stackrel}
\newc{\f}    {\frac}

\begin{document}

\title{ Bremsstrahlung Radiation  At a Vacuum Bubble Wall }
\author{Jae-Weon Lee}\email{scikid@kias.re.kr}
\affiliation{School of Computational Sciences,
             Korea Institute for Advanced Study,
             207-43 Cheongnyangni 2-dong, Dongdaemun-gu, Seoul 130-722, Korea}
\author{ Kyungsub Kim
and Chul H. Lee}
\affiliation{
      Department of Physics,
               Hanyang University,
                Seoul 133-791, Korea}

\author{Ji-ho Jang}
\affiliation{Korea Atomic Energy Research Institute Yuseong, Daejeon 305-353, Korea
}

\begin{abstract}
 When charged particles collide with a vacuum  bubble,
  they can radiate strong
 electromagnetic waves due to rapid deceleration. Owing
to the energy loss of the particles by this bremsstrahlung
 radiation, there is a non-negligible damping
pressure acting on the bubble wall even when  thermal equilibrium
is maintained. In the non-relativistic region, this pressure is proportional to the velocity of the wall
and could have influenced the bubble dynamics in the early universe.
\end{abstract}
\maketitle

There have been many studies on  cosmological roles of
 first-order phase transitions, which proceed by  nucleations and
collisions of vacuum bubbles\cite{chlee1}.
For example, in electroweak baryogenesis models\cite{RevModPhys.71.1463}
rapid bubble expansion can provide a non-equilibrium environment, which may result in
asymmetry between matter and antimatter.
Furthermore, in  some inflationary models\cite{PhysRevLett.62.376,PhysRevLett.67.3639,koh-2005}, the speed of
expanding vacuum bubbles determines how long the inflation period  lasts.
 To understand the bubble kinematics
in a hot plasma, it is important to study  particle scatterings at
a moving bubble wall. To calculate the velocity of
electro-weak
bubbles\cite{PhysRevLett.68.1803,PhysRevLett.75.777,PhysRevD.25.2074,PhysRevD.45.3415,PhysRevD.46.550,chlee2}
and the CP violating charge transport rate  by the wall available
for baryogenesis\cite{RevModPhys.71.1463}, one should know the
reaction force acting on the wall due to the scattered particles, such
as  quarks and gauge bosons\cite{kaplan1,PhysRevD.50.774}.
(For a supersymmetric model see, for example, Ref. \citealp{john-2003-648})

At the first order cosmological phase transition, the false vacuum
 decays to the true vacuum, which has lower energy, by
 making a vacuum bubble.
 When it is created, the wall of the bubble is at rest. As the free energy difference
 between the inner and the outer parts of the bubble fuels the wall, the velocity of the wall
 increases to the light velocity unless there is a damping force.
In the literature, it is generally believed that the non-trivial damping force is
caused by a
deviation of the particle population from a thermal equilibrium one.
In this paper, we study the effect of bremsstrahlung radiations emitted by  particles
on the pressure acting on a bubble wall (not necessary  electroweak bubbles)
during cosmological first order phase transitions.

The aim of this work is to show that, contrary to the usual arguments,  the radiation damping
could give a non-negligible  pressure even when the particles maintain  thermal equilibrium.
Bremsstrahlung (braking radiation) is a radiation due to the
acceleration or deceleration of a charged particle\cite{Byun}.
Entering a true vacuum through a bubble wall, particles interact with the
wall and  could acquire mass and be
decelerated. For example, a fermion field $\psi$ can get  mass through the well-known
Yukawa term $g\bar\psi\phi\psi=m \bar\psi \psi$, where $\phi$ is a Higgs field.
At this time, if the particle is charged electromagnetically,
it can radiate strong electromagnetic waves due to the deceleration.
Let us calculate the pressure from the scattering.

For simplicity, we assume a linear profile for
the bubble wall, i.e., $g\phi(x)\equiv m(x)=m_0 x/d$ when $0<x<d$.
(See Fig.1.)
and choose the coordinates of the rest frame of the bubble wall. This
approximation is good for the usual $tanh$ profile of the wall. The
radiation power of an accelerated particle is given by a
relativistic version of the Larmor's formula\cite{Jackson}:
\beq
\frac{dE_{rad}}{dt}=\frac{A}{m^2}\left (\frac{d\vec{k}}{dt}
\right)^2, \label{lamor}
\eeq
 where $A=2e^2/3c^3\simeq0.0611$ in
the natural units ($\hbar=c=k=1$) and $\vec{k}=(k_x,k_y,k_z)$ is the
3-momentum of a particle.
We  assume a situation
where this classical description of  bremsstrahlung is good
enough. Also, assuming that the wall is planar and parallel to the
y-z plane, we can treat the bubble as a 1-dimensional one along the
$x$-axis. The energy, momentum, and mass of the particle satisfy the
usual relation \beq E^2\equiv m^2(x)+\vec{k}^2(x). \eeq Let us
denote the x-component of the momentum ($k_x$) as $k$ from now on.
Differentiating  the above equation with time  $t$ and using
$dx/dt\equiv v$ and $k=Ev$, we get the force acting on
the wall due to the particles \beq
\frac{dk}{dt}=-\frac{dm^2}{dx}\frac{1}{2E}, \label{dkdt} \eeq
which is the starting point of the pressure
calculation\cite{PhysRevLett.68.1803}. However, if we also
consider the energy carried  away by
 the radiation $E_{rad}$, then the total energy conserved
is $E_{tot}\equiv E+E_{rad}$ and the force  and, hence, the
pressure should be changed. From $dE_{tot}/dt=0$, we obtain \beq
\frac{k}{E}\frac{dk}{dt}+\frac{m}{E}\frac{dm}{dt} +\frac{A}{m^2}
\left( \frac{dk}{dt} \right)^2=0, \eeq which has a solution for
the force \beq \frac{dk}{dt}=\frac{-m^3}{2A E}\left
[1-\sqrt{1-\frac{4A E k} {m^4}\frac{dm}{dt}}\right]. \eeq
Up to
$O(A)$, one can expand the square root term and obtain \beq
\frac{dk}{dt}\simeq -\frac{dm}{dx}\frac{m}{E} - \frac{2A}{kE}\left
(\frac{dm}{dx} \right )^2. \label{dk} \eeq The second term
represents the radiation damping. Then, the total pressure due to the
collisions of the particles in the plasma is given
by\cite{PhysRevLett.68.1803} \beq P=\int^{\infty}_{-\infty}{dx}
\int \frac{d^3\vec{k}}{(2\pi)^3}\left [-\frac{dk}{dt}
f(E(k))\right ], \eeq where $f(E)=(exp(\beta E)\pm 1)^{-1}$ is a
distribution function of  fermions and bosons, respectively.

 First,
let us briefly review the well-known results without radiation
damping. When the mean velocity of the plasma fluid $V$ relative
to the wall (or the negative of the bubble wall velocity relative to the
fluid ) is zero, the first term of  Eq. (\ref{dk}) contributes
\beqa
 P_1&=&\int^{\infty}_{-\infty}\frac{dm^2(x)}{dx} dx \int
\frac{d^3\vec{k}}{(2\pi)^3}\frac{1}{2E}
\frac{1}{e^{\beta E}  \pm 1}, \nonumber \\
&=& F(m_0,T)-F(0,T),
\eeqa
where $F(\phi,T)$ is a free energy of
$\phi$ at a temperature $T=\beta^{-1}$. When $V\neq 0$, the
distribution function is changed to \beq
f[\gamma(E-Vk)]=\left(e^{\beta \gamma(E-Vk)}\pm 1 \right)^{-1}.
\eeq
Here, $\gamma=(1-V^2)^{-\frac{1}{2}}$.
However, using the fact that the phase factor $d^3\vec{k}/E$
is a Lorentz invariant and changing the integration variable to
$k'=\gamma (k-VE)$ and defining $E'\equiv \gamma (E-Vk)$, one can
find that the $V$ dependency of $P_1$ disappears
\cite{PhysRevLett.68.1803}. From this, it is generally believed
that to get non-trivial pressure on the wall, one  needs to consider
a non-equilibrium deviation of $f$\cite{PhysRevLett.75.777}. Our
work indicates this is not necessarily true for some phase
transitions. To see this, consider the effect of the radiation (the
second term of Eq. (\ref{dk})). When $V=0$, the term contributes to
the pressure \beq P_2=2A\int^{\infty}_{-\infty}\left(
\frac{dm(x)}{dx} \right )^2 dx
\int \frac{d^3k}{(2\pi)^3}\frac{1}{Ek(e^{\beta E}  \pm 1)}, \nonumber \\
\eeq
which also vanishes because the second integrand is an odd function of k.
However, when $V\neq 0$,
one can easily check that,
due to the $1/k$ term, the $V$ dependency survives
even under the change of the integration variable.
Thus, in this case,
\beqa
P_2 &=& 2A\int^{\infty}_{-\infty}\left( \frac{dm(x)}{dx} \right
)^2 dx
\int \frac{d^3\vec{k}}{(2\pi)^3}\frac{1}{Ek(e^{\beta \gamma(E-Vk)} \pm 1)}. \nonumber \\
&\equiv&2 A\int^{\infty}_{-\infty}\left( \frac{dm(x)}{dx} \right
)^2 dx ~ I_2(x).
\label{p2}
\eeqa

To be more concrete, let us calculate an approximate value
of the integration when $V\ll1$ for fermions.
In this case, we can expand $f[\gamma(E-Vk)]\simeq f(E)-V\beta k f(E)[f(E)-1]
=f(E)+V\beta k f^{2}(E) exp(-\beta E)$.
The integration of the first term gives zero,
and the second term contributes
\beqa
I_2=V\beta \int \frac{d^3\vec{k}}{(2\pi)^3}\frac{1}{E}f^{2}(E)
exp(-\beta E)\simeq\frac{(ln 2) T V}{2\pi^2}
\eeqa
 because
\beq
\int \frac{d^3k}{(2\pi)^3}\frac{1}{E}f^{2}(E)
exp(-\beta E)\simeq \frac{(ln2) T^2}{2\pi^2},
\eeq
 to lowest oder in $(m/T)^2$ (See Ref. \citealp{PhysRevLett.75.777}).
Therefore, for the wall described in Fig. 1 the pressure by the
radiation is \beq
\label{p2approx} P_2\simeq \frac{(ln 2)A m_0^2 T
}{\pi^2 d} V,
 \eeq
  which is comparable to the result of numerical integration
of Eq. (\ref{p2}) for $V\ll 1$, as shown in Fig. 2. (During the numerical study
it is useful to change the measure from $dk_ydk_z$ to $2\pi E dE$.)
This pressure is proportional to the wall velocity
 up to the moderately relativistic case
  and exists even when  the system is in a thermal equilibrium.
 During the electroweak phase transition, a particle's electromagnetic charge
is not definite, so the $A$ value in  Lamor's formula can not be a constant. In this paper
, however, to perform
a rough calculation, we have assumed that  A is a constant during the phase transition.
For illustration of high temperature effects on electric charges, now we consider a Debye screening of electric charge by plasma
during the phase transition, which is given by effective coupling
$\alpha_{eff}=\alpha/(1-2\alpha ~ln(k/\Lambda)/3\pi)\simeq 0.97\alpha$, where we used
averaged momentum $\langle k \rangle\simeq 3 T$ and $\Lambda$ of order electron mass
at the last approximation
 (see Eq. (42) of ~\cite{Schneider:2002jc}). Thus, we obtain $A=0.0599$ which is
slightly smaller than the zero temperature value.
We also plot the pressure with this $A$ value.

It is noteworthy that the pressure caused by the radiation damping (Eq.
(\ref{p2approx})) is of order $O(\alpha)$, which is bigger than
the pressure due to a departure from thermal
equilibriums\cite{moore-2000-0003,PhysRevD.52.7182,PhysRevLett.75.777}
($O(\alpha^2)$)\cite{moore-2000-0003}, and hence
non-negligible. Here, $\alpha$ is the fine structure constant. Note
also that the power of bremsstrahlung due to bubble walls is much
stronger ($O(\alpha)$) than
 that of ordinary bremsstrahlung of electrons colliding with ions in a plasma ($O(\alpha^3)$)\cite{Ichimaru}.
Since the electroweak phase transition is a complicated phenomenon, by no means is our work
 a full  calculation of the  pressure acting on the electroweak bubbles.
The purpose of this paper  is to present a general idea   that
radiation damping (although usually ignored in
the many related works for bubble wall velocity calculations) could give rise to
significant frictional forces even in thermal
equilibrium states at some cosmological phase transitions.
To include the effects of other particles (e.g., gluon and
W/Z particles) in our work, we need to modify Larmor's formula
by using some sort of group factor. Even in this case, it is hardly probable
that the pressure from the radiation damping from different gauge sectors exactly cancel each other.
Hence, one can expect that a $O(\alpha)$ viscosity  to survive.
Since  bubbles are slow initially, they are supposed to be in a thermal equilibrium state initially.
An ordinary calculation shows no friction at this time, but the radiation damping force exists in this stage,
and  hence, this pressure can significantly change the early evolution of the vacuum bubbles, and
 the nature of electroweak baryogenesis or inflationary cosmology.

\vskip 5.4mm
\section*{ACKNOWLEDGEMENTS}
The authors are thankful to  Myongtak Choi for
helpful discussions.
This work was supported in part by the Korean Science and Engineering Foundation
and Korea Research
Foundation (BSRI-98-2441).

\begin{thebibliography}{20}
\expandafter\ifx\csname natexlab\endcsname\relax\def\natexlab#1{#1}\fi
\expandafter\ifx\csname bibnamefont\endcsname\relax
  \def\bibnamefont#1{#1}\fi
\expandafter\ifx\csname bibfnamefont\endcsname\relax
  \def\bibfnamefont#1{#1}\fi
\expandafter\ifx\csname citenamefont\endcsname\relax
  \def\citenamefont#1{#1}\fi
\expandafter\ifx\csname url\endcsname\relax
  \def\url#1{\texttt{#1}}\fi
\expandafter\ifx\csname urlprefix\endcsname\relax\def\urlprefix{URL }\fi
\providecommand{\bibinfo}[2]{#2}
\providecommand{\eprint}[2][]{\url{#2}}

\bibitem[{\citenamefont{Lee}(1998{\natexlab{a}})}]{chlee1}
\bibinfo{author}{\bibfnamefont{C.~H.} \bibnamefont{Lee}},
  \bibinfo{journal}{J. Korean Phys. Soc.} \textbf{\bibinfo{volume}{33}}, \bibinfo{pages}{588}
  (\bibinfo{year}{1998}{\natexlab{a}}).

\bibitem[{\citenamefont{Trodden}(1999)}]{RevModPhys.71.1463}
\bibinfo{author}{\bibfnamefont{M.}~\bibnamefont{Trodden}},
  \bibinfo{journal}{Rev. Mod. Phys.} \textbf{\bibinfo{volume}{71}},
  \bibinfo{pages}{1463} (\bibinfo{year}{1999}).

\bibitem[{\citenamefont{La and Steinhardt}(1989)}]{PhysRevLett.62.376}
\bibinfo{author}{\bibfnamefont{D.}~\bibnamefont{La}} \bibnamefont{and}
  \bibinfo{author}{\bibfnamefont{P.~J.} \bibnamefont{Steinhardt}},
  \bibinfo{journal}{Phys. Rev. Lett.} \textbf{\bibinfo{volume}{62}},
  \bibinfo{pages}{376} (\bibinfo{year}{1989}).

\bibitem[{\citenamefont{Goldwirth and Zaglauer}(1991)}]{PhysRevLett.67.3639}
\bibinfo{author}{\bibfnamefont{D.~S.} \bibnamefont{Goldwirth}}
  \bibnamefont{and} \bibinfo{author}{\bibfnamefont{H.~W.}
  \bibnamefont{Zaglauer}}, \bibinfo{journal}{Phys. Rev. Lett.}
  \textbf{\bibinfo{volume}{67}}, \bibinfo{pages}{3639} (\bibinfo{year}{1991}).

\bibitem[{\citenamefont{Koh}(2005)}]{koh-2005}
\bibinfo{author}{\bibfnamefont{S.}~\bibnamefont{Koh}}, \bibinfo{journal}{J. Korean Phys. Soc.}
  \textbf{\bibinfo{volume}{49}}, \bibinfo{pages}{787} (\bibinfo{year}{2005}).

\bibitem[{\citenamefont{Turok}(1992)}]{PhysRevLett.68.1803}
\bibinfo{author}{\bibfnamefont{N.}~\bibnamefont{Turok}},
  \bibinfo{journal}{Phys. Rev. Lett.} \textbf{\bibinfo{volume}{68}},
  \bibinfo{pages}{1803} (\bibinfo{year}{1992}).

\bibitem[{\citenamefont{Moore and
  Prokopec}(1995{\natexlab{a}})}]{PhysRevLett.75.777}
\bibinfo{author}{\bibfnamefont{G.}~\bibnamefont{Moore}} \bibnamefont{and}
  \bibinfo{author}{\bibfnamefont{T.}~\bibnamefont{Prokopec}},
  \bibinfo{journal}{Phys. Rev. Lett.} \textbf{\bibinfo{volume}{75}},
  \bibinfo{pages}{777} (\bibinfo{year}{1995}{\natexlab{a}}).

\bibitem[{\citenamefont{Steinhardt}(1982)}]{PhysRevD.25.2074}
\bibinfo{author}{\bibfnamefont{P.~J.} \bibnamefont{Steinhardt}},
  \bibinfo{journal}{Phys. Rev. D} \textbf{\bibinfo{volume}{25}},
  \bibinfo{pages}{2074} (\bibinfo{year}{1982}).

\bibitem[{\citenamefont{Enqvist et~al.}(1992)\citenamefont{Enqvist, Ignatius,
  Kajantie, and Rummukainen}}]{PhysRevD.45.3415}
\bibinfo{author}{\bibfnamefont{K.}~\bibnamefont{Enqvist}},
  \bibinfo{author}{\bibfnamefont{J.}~\bibnamefont{Ignatius}},
  \bibinfo{author}{\bibfnamefont{K.}~\bibnamefont{Kajantie}}, \bibnamefont{and}
  \bibinfo{author}{\bibfnamefont{K.}~\bibnamefont{Rummukainen}},
  \bibinfo{journal}{Phys. Rev. D} \textbf{\bibinfo{volume}{45}},
  \bibinfo{pages}{3415} (\bibinfo{year}{1992}).

\bibitem[{\citenamefont{Dine et~al.}(1992)\citenamefont{Dine, Leigh, Huet,
  Linde, and Linde}}]{PhysRevD.46.550}
\bibinfo{author}{\bibfnamefont{M.}~\bibnamefont{Dine}},
  \bibinfo{author}{\bibfnamefont{R.~G.} \bibnamefont{Leigh}},
  \bibinfo{author}{\bibfnamefont{P.}~\bibnamefont{Huet}},
  \bibinfo{author}{\bibfnamefont{A.}~\bibnamefont{Linde}}, \bibnamefont{and}
  \bibinfo{author}{\bibfnamefont{D.}~\bibnamefont{Linde}},
  \bibinfo{journal}{Phys. Rev. D} \textbf{\bibinfo{volume}{46}},
  \bibinfo{pages}{550} (\bibinfo{year}{1992}).

\bibitem[{\citenamefont{Lee}(1998{\natexlab{b}})}]{chlee2}
\bibinfo{author}{\bibfnamefont{C.~H.} \bibnamefont{Lee}},
  \bibinfo{journal}{J. Korean Phys. Soc.} \textbf{\bibinfo{volume}{32}}, \bibinfo{pages}{861}
  (\bibinfo{year}{1998}{\natexlab{b}}).

\bibitem[{\citenamefont{Andrew G.~Cohen and Nelson}(1991)}]{kaplan1}
\bibinfo{author}{\bibfnamefont{D.~B.~K.} \bibnamefont{Andrew G.~Cohen}}
  \bibnamefont{and} \bibinfo{author}{\bibfnamefont{A.~E.}
  \bibnamefont{Nelson}}, \bibinfo{journal}{Nuc. Phys. B}
  \textbf{\bibinfo{volume}{349}}, \bibinfo{pages}{727} (\bibinfo{year}{1991}).

\bibitem[{\citenamefont{Farrar and Shaposhnikov}(1994)}]{PhysRevD.50.774}
\bibinfo{author}{\bibfnamefont{G.~R.} \bibnamefont{Farrar}} \bibnamefont{and}
  \bibinfo{author}{\bibfnamefont{M.~E.} \bibnamefont{Shaposhnikov}},
  \bibinfo{journal}{Phys. Rev. D} \textbf{\bibinfo{volume}{50}},
  \bibinfo{pages}{774} (\bibinfo{year}{1994}).

\bibitem[{\citenamefont{John and Schmidt}(2001)}]{john-2003-648}
\bibinfo{author}{\bibfnamefont{P.}~\bibnamefont{John}} \bibnamefont{and}
  \bibinfo{author}{\bibfnamefont{M.~G.} \bibnamefont{Schmidt}},
  \bibinfo{journal}{Nucl. Phys. B} \textbf{\bibinfo{volume}{598}},
  \bibinfo{pages}{291} (\bibinfo{year}{2001}).

\bibitem[{\citenamefont{Byun et~al.}(2005)\citenamefont{Byun, Kim, and
  Kwak}}]{Byun}
\bibinfo{author}{\bibfnamefont{K. T.} \bibnamefont{Byun}},
  \bibinfo{author}{\bibfnamefont{K. Y.} \bibnamefont{Kim}}, \bibnamefont{and}
  \bibinfo{author}{\bibfnamefont{H. Y.} \bibnamefont{Kwak}},
  \bibinfo{journal}{J. Korean Phys. Soc. } \textbf{\bibinfo{volume}{47}}, \bibinfo{pages}{1010}
  (\bibinfo{year}{2005}).

\bibitem[{\citenamefont{Jackson}(1975)}]{Jackson}
\bibinfo{author}{\bibfnamefont{J.}~\bibnamefont{Jackson}},
  \emph{\bibinfo{title}{Classical Electrodynamics, 2nd ed.}}
  (\bibinfo{publisher}{Wiley, New York}, \bibinfo{year}{1975}).

\bibitem[{\citenamefont{Schneider}(2002)}]{Schneider:2002jc}
\bibinfo{author}{\bibfnamefont{R.~A.} \bibnamefont{Schneider}},
  \bibinfo{journal}{Phys. Rev.} \textbf{\bibinfo{volume}{D66}},
  \bibinfo{pages}{036003} (\bibinfo{year}{2002}).

\bibitem[{\citenamefont{Moore}(2000)}]{moore-2000-0003}
\bibinfo{author}{\bibfnamefont{G.~D.} \bibnamefont{Moore}},
  \bibinfo{journal}{JHEP} \textbf{\bibinfo{volume}{0003}}, \bibinfo{pages}{006}
  (\bibinfo{year}{2000}).

\bibitem[{\citenamefont{Moore and
  Prokopec}(1995{\natexlab{b}})}]{PhysRevD.52.7182}
\bibinfo{author}{\bibfnamefont{G.~D.} \bibnamefont{Moore}} \bibnamefont{and}
  \bibinfo{author}{\bibfnamefont{T.}~\bibnamefont{Prokopec}},
  \bibinfo{journal}{Phys. Rev. D} \textbf{\bibinfo{volume}{52}},
  \bibinfo{pages}{7182} (\bibinfo{year}{1995}{\natexlab{b}}).

\bibitem[{\citenamefont{Ichimaru}(1973)}]{Ichimaru}
\bibinfo{author}{\bibfnamefont{S.}~\bibnamefont{Ichimaru}},
  \emph{\bibinfo{title}{Basic Principles of Plasma Physics}}
  (\bibinfo{publisher}{W. A. Benjamin, Reading, MA.}, \bibinfo{year}{1973}).

\end{thebibliography}

\newpage

\begin{figure}[htbp]  \label{fig1}
\includegraphics[width=\textwidth]{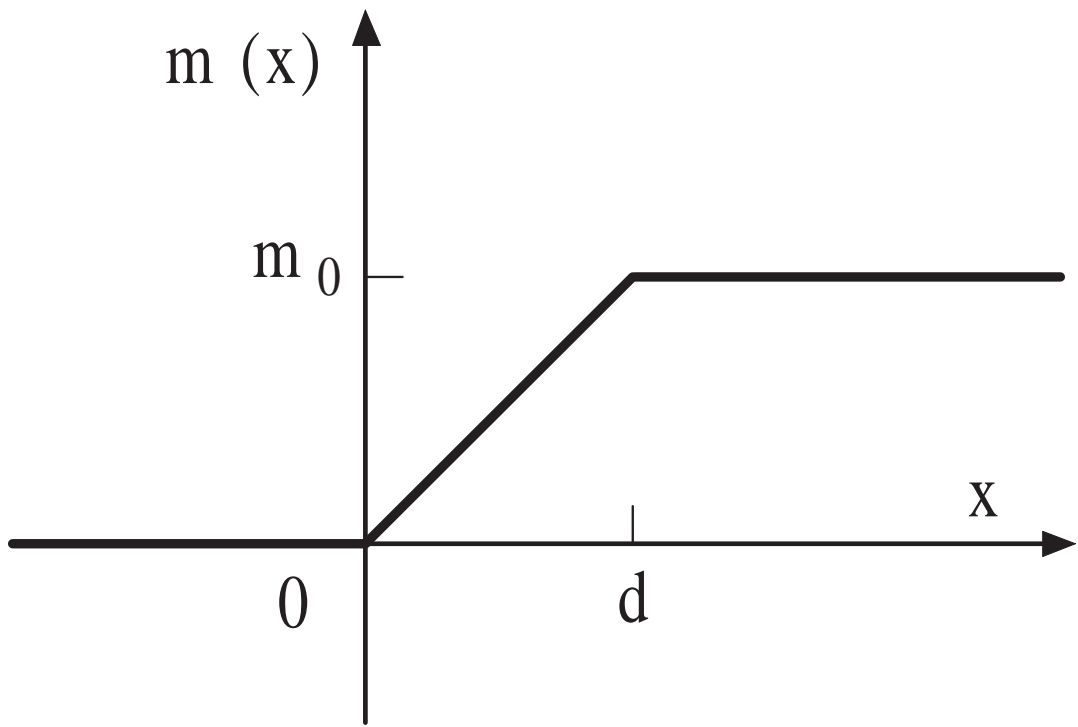}
\caption{ The effective mass of the particle $m(x)$ in the wall rest frame.
 }
\end{figure}

\begin{figure}[htbp]  \label{fig2}
\includegraphics[width=\textwidth]{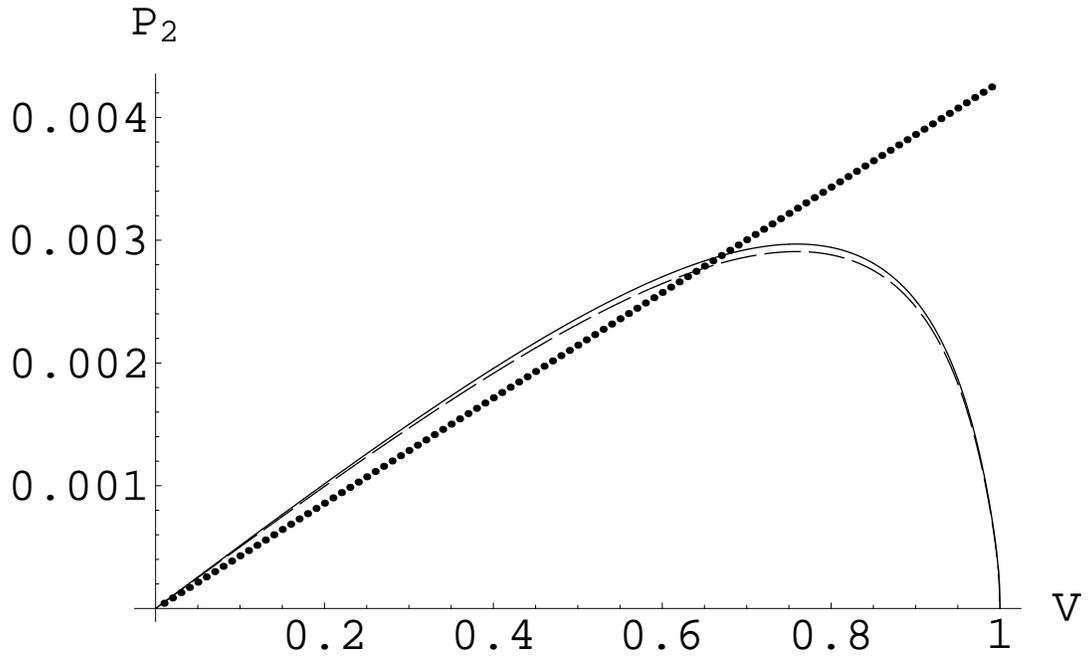}
\caption{  The pressure by the radiation damping of fermions
colliding with the linear bubble wall  as a function of the wall velocity.
The thick line shows numerical integration of Eq. (\ref{p2}) and the
dotted line shows the approximate formula in Eq. (\ref{p2approx}).
 Here we set
$1/d=1=m_0=T$ for simplicity.
The dashed line represents the result with Debye screening of charge. }
\end{figure}

\end{document}